\documentclass[twocolumn,twoside,
  letterpaper,superscriptaddress,amsmath,amssymb,pra]{revtex4}
\usepackage[dvips]{graphicx}
\usepackage[dvips,a4paper]{geometry}
\usepackage{amssymb}
\usepackage{dcolumn}
\usepackage{bm}
\usepackage{verbatim}
\usepackage{eucal}
\usepackage{color}
\definecolor{DarkGreen}{rgb}{0.00,0.39,0.00}

\newcommand{\ket}[1]{\ensuremath{\left| #1 \right\rangle}}

\newcommand {\eq}{\begin{equation}}
\newcommand {\eeq}{\end{equation}}
\newcommand {\eqa}{\begin{eqnarray}}
\newcommand {\eeqa}{\end{eqnarray}}

\begin{document}

\title{High-fidelity transmission of polarization encoded qubits from an entangled source over 100~km of fiber}




\newcommand{\Faculty}{Quantum Optics, Quantum Nanophysics and
  Quantum Information, Faculty of Physics, University of Vienna,
  Boltzmanngasse~5, 1090 Vienna, Austria}
\newcommand{\ARC}{Austrian Research Centers~GmbH~-~ARC,
  Donau-City-Str.~1, 1220 Vienna, Austria}  
\newcommand{\Institute}{Institute for Quantum Optics and Quantum Information,
Austrian Academy of Sciences, Boltzmanngasse 3, 1090 Vienna,
Austria}

\author{Hannes H\"{u}bel}
\email[Corresponding author:~]{hannes.huebel@univie.ac.at}
\affiliation{\Faculty}
\author{Michael R. Vanner}
\affiliation{\Faculty}
\author{Thomas Lederer}
\affiliation{\Faculty}
\author{Bibiane~Blauensteiner}
\affiliation{\Faculty}
\author{Thomas Lor\"{u}nser}
\affiliation{\ARC}
\author{Andreas Poppe}
\affiliation{\Faculty}
\author{Anton Zeilinger}
\affiliation{\Faculty}\affiliation{\Institute}

\date{June the 8th, 2007}




\begin{abstract}
We demonstrate non-degenerate down-conversion at 810 and 1550~nm for
long-distance fiber based quantum communication using polarization
entangled photon pairs. Measurements of the two-photon visibility,
without dark count subtraction, have shown that the quantum
correlations (raw visibility 89\%) allow secure quantum cryptography
after 100~km of non-zero dispersion shifted fiber using commercially
available single photon detectors. In addition, quantum state
tomography has revealed little degradation of state negativity,
decreasing from 0.99 at the source to 0.93 after 100~km, indicating
minimal loss in fidelity during the transmission.
\end{abstract}


\maketitle

\section{Introduction}
The use of quantum entanglement in combination with existing fiber
telecommunication networks offers the possibility to implement
long-distance quantum communication protocols like quantum key
distribution (QKD)~\cite{cryptodusek, BBM92} and applications such
as quantum repeaters~\cite{briegel} in the future. If the existing
infrastructure is used several prerequisites have to be met:
Firstly, the entanglement has to be shared by at least one photon in
the 1550~nm telecom band to utilize low fiber absorption. Secondly,
the fidelity of the quantum correlations after transmission has to
be sufficient for the detection of eavesdropping without performing
post-corrections and thirdly, the detection rate of photon pairs has
to be high enough for a meaningful application and also to overcome
the limitation of detector dark counts.

In this work we report a high generation rate of polarization
entangled photon pairs at 810 and 1550~nm by asymmetric spontaneous
parametric down-conversion (SPDC) and investigations into any
depolarization of the 1550~nm photon with fiber transmission by
means of measurement of the two-photon visibility and quantum state
tomography. This is the first time that quantum correlations have
been distributed over 100~km of optical fiber sufficiently high for
QKD without any post-corrections. In the QKD sheme, it is essential
to attribute all measured errors to the eavesdropper and hence no
background subtraction is allowed. In addition, a high detection
rate of $\sim$100 pairs/s after 100~km, makes our distribution
scheme readily practical for long-distance quantum communication
applications.

Using the polarization degree of freedom of single photons as a
carrier of information has long been seen as disadvantageous in
fiber transmission. Although long-distance transmission of
polarization encoded qubits was successfully demonstrated in free
space up to 144 km~\cite{freeSpace3}, birefringent effects of the
fiber were believed to hinder long-distance transmission of the
polarization state due to depolarization, mainly polarization mode
dispersion (PMD). With these effects in mind, long-distance fiber
transmission has been performed using time-bin entangled photons,
reaching distances of 50~km~\cite{Marcikic04} and
60~km~\cite{Takesue06}.

Recent experiments have demonstrated polarization entanglement
distribution up to 100~km~\cite{Kumar100km} in fibers. However, in
these experiments quantum correlation could only be shown after
subtraction of background counts. In addition, the use of standard
wavelength-division-multiplexing components in the quantum channel
was demonstrated for polarization entangled photon
pairs~\cite{Kumar100km,sebo} combining classical and quantum signals
in a single fiber.

These results are encouraging as polarization encoded qubits can be
easily produced, manipulated and detected with simple optical
components. Hence the use of polarization entanglement in optical
fibers brings several advantages; for instance, the construction of
entanglement based QKD setups with only passive components reduces
the risk of side channel attacks. Also, it allows more convenient
control for interaction with atoms, thereby giving greater
opportunity for the development of quantum memory and repeater
devices (e.g. conversion of polarization encoded photons into atomic
qubits~\cite{photonatom}).

\begin{figure}
\begin{center}
\includegraphics[width=\linewidth]{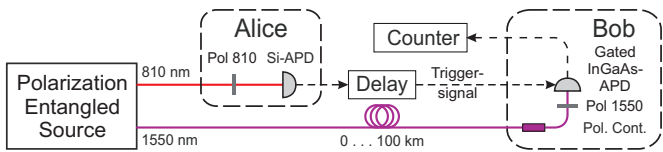}
\caption{(color online) Distribution and Measurement Scheme: A
source of polarization entangled photon pairs produces photons at
810 and 1550~nm. The 810~nm pair-photons are polarization analyzed
locally by Alice with a rotatable polarizer (Pol 810) and
subsequently detected by her silicon APD (Si-APD). The partner
photons at 1550~nm are transmitted to Bob via telecom fibers on
spools, analyzed (Pol 1550) and detected by an InGaAs-APD. The
random rotation in polarization is compensated  by an electronic
polarization controller (Pol.Cont.). The trigger signals are
carefully matched by an electronic delay generator (Delay) to the
transmission time through the fiber spools and the measured
coincidence rate is displayed on a counter.} \label{DistScheme}
\end{center}
\end{figure}

Our two-photon distribution and measurement scheme is depicted in
Fig.~\ref{DistScheme}. A source of polarization entanglement
distributes photon pairs to Alice and Bob. Alice performs local
polarization analysis on the 810~nm photon of the pair. The 1550~nm
photon is transmitted to Bob who performs his own polarization
analysis. The measured coincidences obtained by Alice and Bob can be
used to characterize the shared quantum state or obtain a secret key
according to the BBM92 protocol for entangled photons~\cite{BBM92}.
Such a QKD scheme has been used to demonstrate the successful
distribution of secret keys in an urban environment~\cite{Poppe04}.

Even though one photon of the entangled pair is measured directly,
projecting the state of the other photon, the correlation carried by
this photon is of quantum nature. For this reason, quantum
communication protocols are made possible even if the measurements
are not performed simultaneously. For example, our setup could in
principle be used in a teleportation experiment, where a Bell state
analysis is performed on the 810~nm photon and the to-be-teleported
state. Our 1550~nm photon would then carry the teleported
polarization state through the fiber. The resulting state of such a
teleportation to a far away receiver is determined by the quantum
correlation. The quality of these correlations, measured by quantum
state tomography, are presented in this paper. In addition we
investigate how the quantum correlation evolves in long distance
fiber transmission and what are the principle causes.

\section{Source of polarization entangled photons}
For long-distance fiber communication systems it is essential to
have a high flux of pair generation, which we realized by producing
entangled photons using spontaneous parametric down-conversion
(SPDC)~\cite{ribordy,karlsson,konigsource} in the orthogonally
oriented two crystal geometry~\cite{Kwiat99}.

\begin{figure}[t]
\begin{center}
\includegraphics[width=\linewidth]{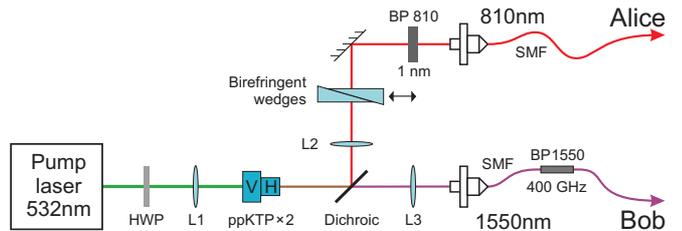}
\caption{(color online) Optical Setup: The solid state diode pumped
laser (Pump laser 532 nm) is focused (L1) at the interface of the
two periodically-poled KTP nonlinear crystals (ppKTP) for highly
non-degenerate collinear down-conversion. The half-wave-plate (HWP)
rotates the pump to excite the horizontal (H) and vertical (V)
crystals equally. The 810 and 1550 nm photons are spatially
separated (Dichroic), recollimated by lens L2 and L3 and coupled
into single-mode fibers (SMF). The wavepackets are spectrally
confined and matched with filters BP 1550 and BP 810. A betraying
timing offset between the photons of one pair is compensated by the
birefringent wedges in the 810 nm arm.} \label{SourceFig}
\end{center}
\end{figure}

Our compact source (40$\times$40~cm), Fig. \ref{SourceFig}, is
pumped by a 532-nm-laser. For equal crystal excitation, the pump
polarization is rotated to $45^{\circ}$ with respect to the crystal
axes and the beam is focussed at the boundary of two crystals. It is
believed that optimal focussing is obtained when the Rayleigh range
$z_0$, is comparable to the individual crystal length $L$ ($L=4$~mm
for both crystals)~\cite{LjunggrenFocus}. The beam size at the focus
was measured to be 55~$\mu$m, giving a Rayleigh range of $\sim$
4.5~mm and hence a $L/z_0$ ratio of 0.9.

The two nonlinear crystals used in the source are quasi-phase
matched periodically-poled KTiOPO$_4$ (ppKTP), with a grating
spacing of 9.7~$\mu$m, which has been tailored for type-I collinear
generation of an asymmetric photon pair at 810 and 1550~nm from a
532~nm pump. These wavelengths were selected because of the
efficient detection at 810~nm, low fiber absorption at 1550~nm and
readily available stable radiation sources at 532~nm. The crystals
are housed in a temperature controlled copper mount, heated to
approximately $65\,^{\circ}\mathrm{C}$. Varying the temperature
allows wavelength tuning for collinear emission. Each ppKTP crystal
produces a pair with an intrinsic bandwidth of 800~GHz, which was
reduced to 400~GHz with filters BP 810 and BP 1550 to minimize
chromatic dispersion.

If the down-conversion processes in the two crystals with orthogonal
polarizations are indistinguishable in terms of spectral, spatial
and temporal degrees of freedom, the presence of a photon pair does
not reveal in which crystal it was produced. The superposition of
the two possible creation events gives rise to the polarization
entangled state of the photon pair:
$\ket{\phi}=\frac{1}{\sqrt{2}}\left(\ket{\rm{H_{810} \hspace{1mm}
H_{1550}}}+e^{i\phi}\ket{\rm{V_{810} \hspace{1mm} V_{1550}}}\right)$

However, chromatic dispersion between the 810 and 1550~nm photon
inside the crystal leads to a temporal distinguishability between
pairs generated in the two different crystals and hence loss of
entanglement~\cite{LjunggrenSource}. This was compensated by
transmission through birefringent quartz wedges which also allow
control of the phase in the entangled state to obtain either of the
two type-I Bell states $\ket{\Phi^\pm}$.

Alice's photons are detected after 2~m of SMF fiber using a silicon
avalanche photo diode (Si-APD), see Fig. \ref{DistScheme}, with an
efficiency of $\sim$50\%. An electronic trigger is sent from the
Si-APD to gate an InGaAs detector (idQuantique id-200) to detect the
corresponding photon at 1550~nm as a coincidence count on Bob's
side. The quantum efficiency ($QE$) at this wavelength is in the
range of 10-15\%. Locally, with 32 m SMF fiber to connect Bob, we
achieved the following rates of polarization entangled photon pairs:

The photon rate at Alice was around 100~kcounts/s/mW. On Bob's side,
with $10~\mu$s of deadtime to reduce afterpulsing, 3000~counts/s/mW
of coincidences were detected in either polarization modes, yielding
a conditional detection probability of $\sim$3\% (ratio of
coincidence to trigger counts). For a low pump power (2~mW) the
uncorrected two-photon visibility was measured to be 99\%. The
visibility is a measure of the polarization correlation between the
photons, a proper definition is given in section 4.2. At maximum
pump power of $\sim$16~mW the measured coincidences were limited to
21000~counts/s mainly due to saturation effects in the InGaAs
detector (limited gate rate of 4~MHz). The degree of entanglement
decreases at higher power levels due to multi pair emission
(production of uncorrelated pairs within the gate time) leading to a
reduced visibility of 97\%. All subsequent measurements were
performed at the maximum power level.

\section{Long-distance fiber transmission}
To determine the robustness of polarization correlations in
long-distance fiber transmission we used the arrangement depicted in
Fig.~\ref{DistScheme}. Since detector dark counts are the limiting
factor in long-distance experiments, the detector bias was reduced
to optimize the signal to noise ratio as represented by
$\frac{QE}{dark counts}$. An optimum was found for a reduced
overbias voltage corresponding to $QE=6\%$ with a dark count rate of
$\sim 3\times10^{-6}$ within each gate opening time. To further
reduce the influence of dark counts, all measurements were performed
at the smallest possible gate width of 1.5~ns on the InGaAs
detector. It was therefore important to minimize chromatic
dispersion (CD) so that the temporal broadening of the photon
($\tau_{P}$) was smaller than the applied gate window. The temporal
broadening ($\tau_P$) due to chromatic dispersion (CD) is given by:
$\tau_P=\,$CD $\cdot \,\lambda_{1550} \cdot L$, where
$\lambda_{1550}$ is the spectral width of the 1550 photon in nm and
$L$ is the fiber length in km.

For the long-distance measurements we used two types of fiber:
Standard single-mode fibers (telecom standard: ITU-T G.562) with
CD$\,\,=\,\,$18~ps/nm/km and non-zero dispersion shifted (NZDS)
fibers (ITU-T G.565) with CD$\,\,\approx\,\,$5~ps/nm/km. For an
initial $\lambda_{1550}\,=\,3.2$~nm, CD would broaden the photon
wavepacket to $\tau_P = 5$ and 1.5~ns after $L=$100~km for standard
and NZDS fibers, respectively. Standard fibers are therefore
inadequate to be used for long distance transmission. The first set
(set I) of our investigations consisted of NZDS fibers with
individual spool lengths of 6.3 and 12.6~km which could be
concatenated (using FC/PC connectors) to give a total length of
63~km and a $\tau_P$ of 1~ns. To reach longer distances, we added
two spools of standard fiber (12.6~km each) totaling 88.2~km and
resulting in an additional CD spread of 1.5~ns (total $\tau_P$ of
2.5~ns). For measurements over even longer distances we had a second
set (set II), consisting of two 50.4~km spools of NZDS fiber spliced
together giving a total transmission distance of 100.8~km and a
$\tau_P$ of 1.5~ns, the size of the gate window.

To gate the InGaAs detector an electronic delay generator was used
to provide delays up to 500 $\mu$s (for 101~km), where fine tuning
by 0.2~ns increments was provided using the internal delay generator
of the detector. Standard delay generators suffer from the problem
that only one pulse at a time can be processed, resulting in a
decreased processing rate as the delay increases. To handle our MHz
rate for long delays, we used a custom made device
(dotfast-consulting).

Due to the birefringence in the fiber, the random-walk like drift in
polarization over the surface of the Poincar\'{e} sphere must be
compensated by Bob before making his polarization analysis. Temporal
phase changes in the order of $\pi/4$ per hour are typical values in
the laboratory. Using an electronic in-fiber polarization
controller, the coincidence counts were minimized, while the
polarizers at Alice and Bob were set to 0$^{\circ}$(\emph{V}) and
90$^{\circ}$(\emph{H}) respectively, which insured compensation in
the \emph{H}\emph{V} basis. In a second step the birefringent
wedges, see Fig.~\ref{SourceFig}, were adjusted to yield a minimal
coincidence rate at 45$^{\circ}$($+$) and 135$^{\circ}$($-$)
settings and obtain a $\ket{\Phi^+}$ state.

Birefringence in the fiber also causes more severe effects on the
polarization state, in particular polarization mode dispersion
(PMD). PMD is a statistical property of the fiber and can be seen as
a concatenation of lossless birefringent elements~\cite{PMD-Buch}.
Here we limit ourselves to first order PMD, described as a set of
two orthogonal axis with a differential-group delay (DGD) between
them. A light pulse, which is not aligned with one of the axis, will
be partially projected along the fast and slow axis. The pulse is
therefore split into two pulses with separation $\tau_{PMD}$, after
transmission. If $\tau_{PMD}$ is larger than the coherence time of
the pulse ($\tau_{coh}$) then the two components will temporally not
overlap and the initial polarization state is destroyed. If the
situation is reversed the pulse will stay together and no
depolarization is caused. The DGD of a fiber is normally stated in
ps$/\sqrt{km}$ with $\tau_{PMD}=DGD\cdot\sqrt{L}$, $L$ being the
length of the fiber. Since we had no means to measure the value of
$\tau_{PMD}$ directly at the time of our experiment, we relied upon
the manufacturers specifications: For set I the average DGD, of our
spools, was around 0.07~ps/$\sqrt{km}$. Individual values of
$\tau_{PMD}$ for each 6.3~km spool varied from 0.1 to 0.2~ps and
between 0.2 to 0.4~ps for the 12.6~km spools. In set II, the maximum
value is 0.1~ps/$\sqrt{km}$. On has to bear in mind that PMD is a
highly statistical property and fluctuations in the environment can
induce large changes in $\tau_{PMD}$. So the above values can only
be seen as an estimate. Nevertheless, these parameters can be used
to determine the significance of PMD. The 400~GHz bandwidth of the
1550~nm photon corresponds to a coherence time of $\tau_{coh}\approx
1.6$~ps. So for 50~km of fiber, $\tau_{PMD}$ is around 0.6~ps,
smaller but still in the range of $\tau_{coh}$. Therefore, PMD
effects are expected to play a role in long distance transmission.

\section{Evolution of the polarization state during transmission}
\subsection{Quantum state tomography}
In order to quantify the effect of the fiber transmission on the
polarization state, full quantum state tomography was performed at
distances of 0, 25.2, 50.4, 75.6, and 100.8~km. Coincidences were
recorded in 16 different settings
 (\emph{H}\emph{H}, \emph{H}\emph{V},\emph{V}\emph{H},
 \emph{V}\emph{V}, \emph{H}$+$, \emph{V}$+$,
 \emph{H}\emph{R}, \emph{V}\emph{R}, $+$\emph{H},
 $+$\emph{V}, $++$,
 $+$\emph{R},
 \emph{R}\emph{H}, \emph{R}\emph{V}, \emph{R}$+$ and
 \emph{R}\emph{R}, where \emph{R} is the right-circular
  polarization)~\cite{james01} each averaged over 10~s and together with a
  maximum likelihood estimation~\cite{maxlike} the density matrix was evaluated.

\begin{figure}[t]
\begin{center}
\includegraphics[width=0.7\linewidth,trim = 0mm 0mm 0mm 0mm, clip]{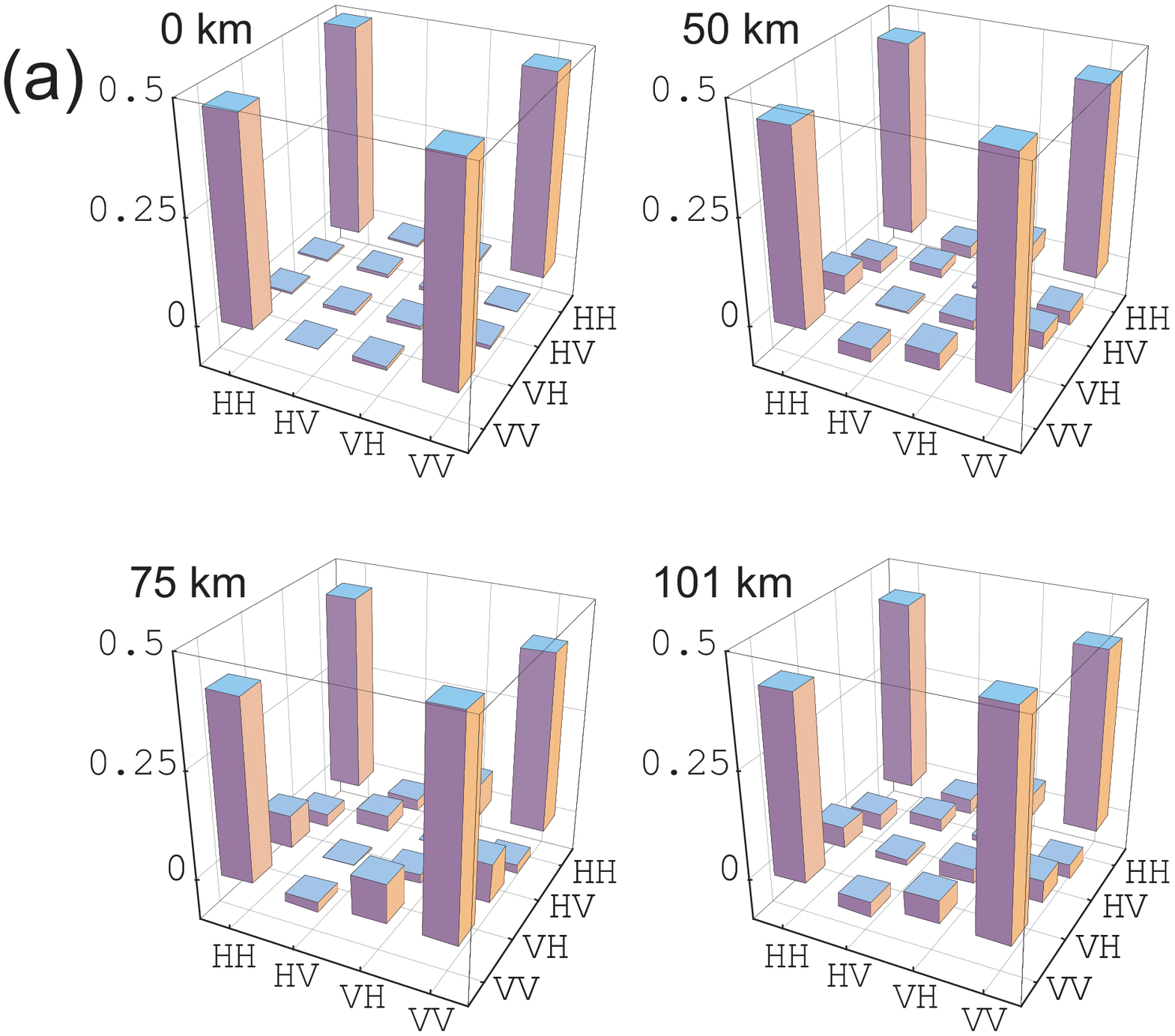}
\hfill
\includegraphics[height=6.1cm,trim = 0mm 0mm 0mm 0mm, clip]{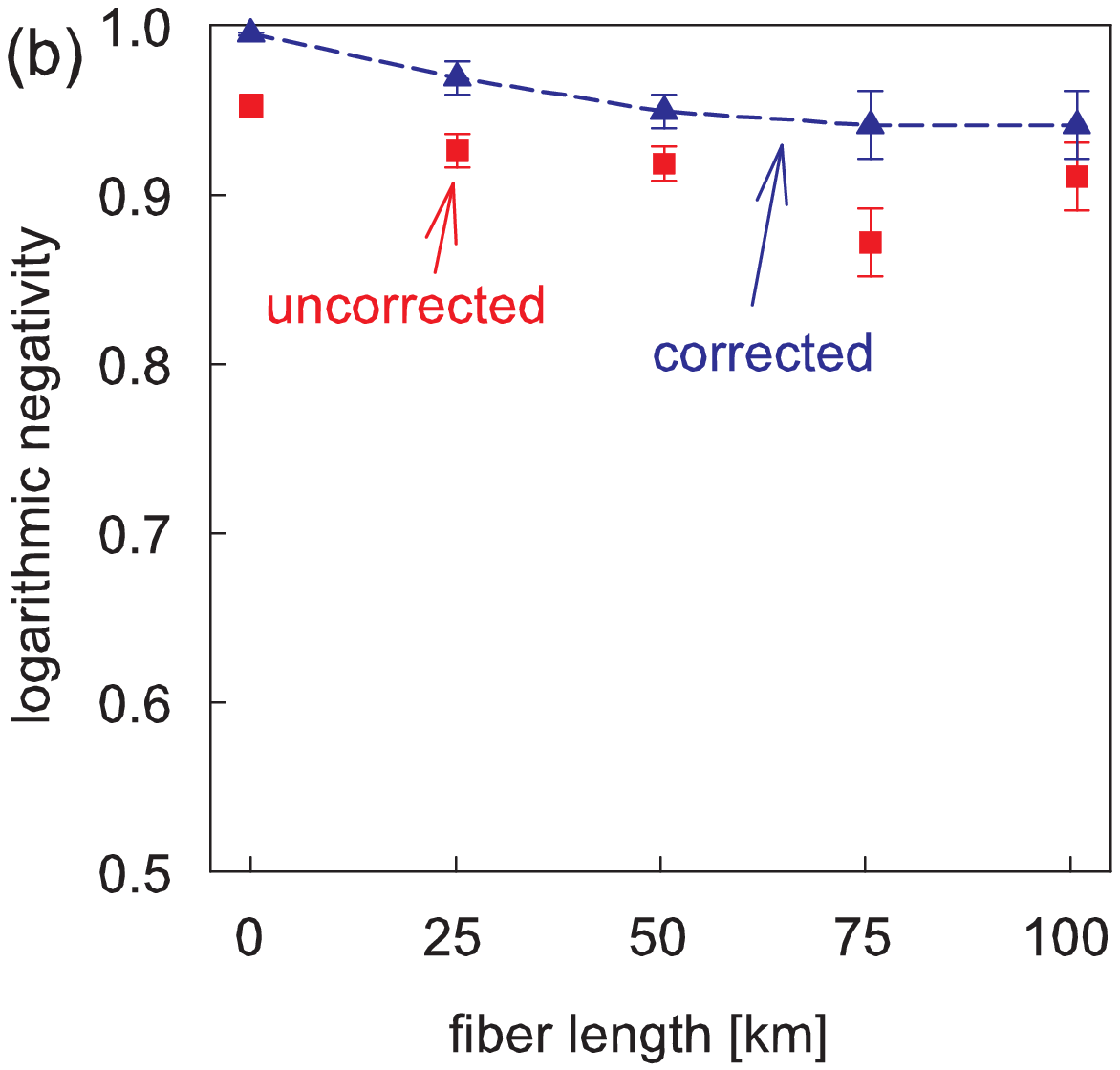}
\caption{(color online) (a) Real part of measured density matrices
at fiber lengths of 0, 50, 75 and 101 km (raw data). The imaginary
part is close to zero with no element higher than 0.09.  (b)
Uncorrected ({\color{red}{\scriptsize$\blacksquare$}}) and corrected
({\color{blue}$\blacktriangle$}) logarithmic negativity. }
\label{tomo}
\end{center}
\end{figure}

The real components of the measured density matrices are depicted in
Fig.~\ref{tomo}(a). The overlap (fidelity) with the pure
$\ket{\Phi^+}$ state (four columns at the corners of the matrix) is
large, even with the matrix obtained at 101~km. To quantitatively
characterize our density matrices and to determine any
depolarization of the state with increasing transmission distance we
chose to use the negativity ($\mathcal{N}$)~\cite{vidal_neg}. The
negativity is a real measure of entanglement by quantifying how
negative the eigenvalues of the density matrix ($\rho$) become after
the partial transpose ($\rho^{T}$) is taken. The higher the
negativity, the more entanglement is present in the system. It is
defined as the absolute value of the sum of negative eigenvalues of
$\rho^{T}$ \eq \mathcal{N}(\rho)= \left|\sum \frac{\mid \lambda_i
\mid - \lambda_i}{2}\right|,\eeq where $\lambda_i$ are the
eigenvalues of $\rho^{T}$.

From $\mathcal{N}$ we can readily compute the logarithmic negativity
$E_\mathcal{N}$ defined as: \eq E_\mathcal{N}(\rho)=
\log_2(2\,\mathcal{N}+1) \eeq

The logarithmic negativity lies between zero (separable states) and
one (maximal entangled states). It also bounds the maximal amount of
entanglement that can be extracted (distilled) from the given state
\cite{destill}.

Figure~\ref{tomo}(b) displays the values of $E_\mathcal{N}$ for
increasing fiber lengths. $E_\mathcal{N}$ is calculated from the raw
data and also from corrected data, where the background has been
removed. The actual background was measured for each individual data
point by adding an additional 10~ns delay thereby temporally
displacing the gate window from the coincidence peak. The recorded
coincidences are then composed of detector dark counts and multi
pair contributions only and were subtracted from the measured
coincidences to calculate a corrected density matrix. The
uncorrected values decrease from 0.94 at 0~km to 0.83 at 75~km. At
101~km, the negativity is higher at 0.88 due to lower losses in set
II. When background counts are removed $E_\mathcal{N}$ decreases
from 0.99 to 0.93 in the first 50~km then remaining constant up to
101~km.

\subsection{Two-photon visibility}
Although the tomography portrays the full information about the
quantum state, in quantum communication applications like QKD only a
subspace of measurement bases is used in practice. We therefore
measured the visibility in the \emph{H}\emph{V} and the $+-$ basis
for fiber lengths up to 101~km. The two-photon visibility ($V$) was
calculated using the definition $V = \frac{Max-Min}{Max+Min}$, where
$Max$ is the maximal coincidence rate as obtained for parallel
polarizer settings at Alice and Bob and $Min$ is the coincidence
rate at orthogonal settings.

\begin{figure}[t]
\begin{center}
\includegraphics[height=6.0cm]{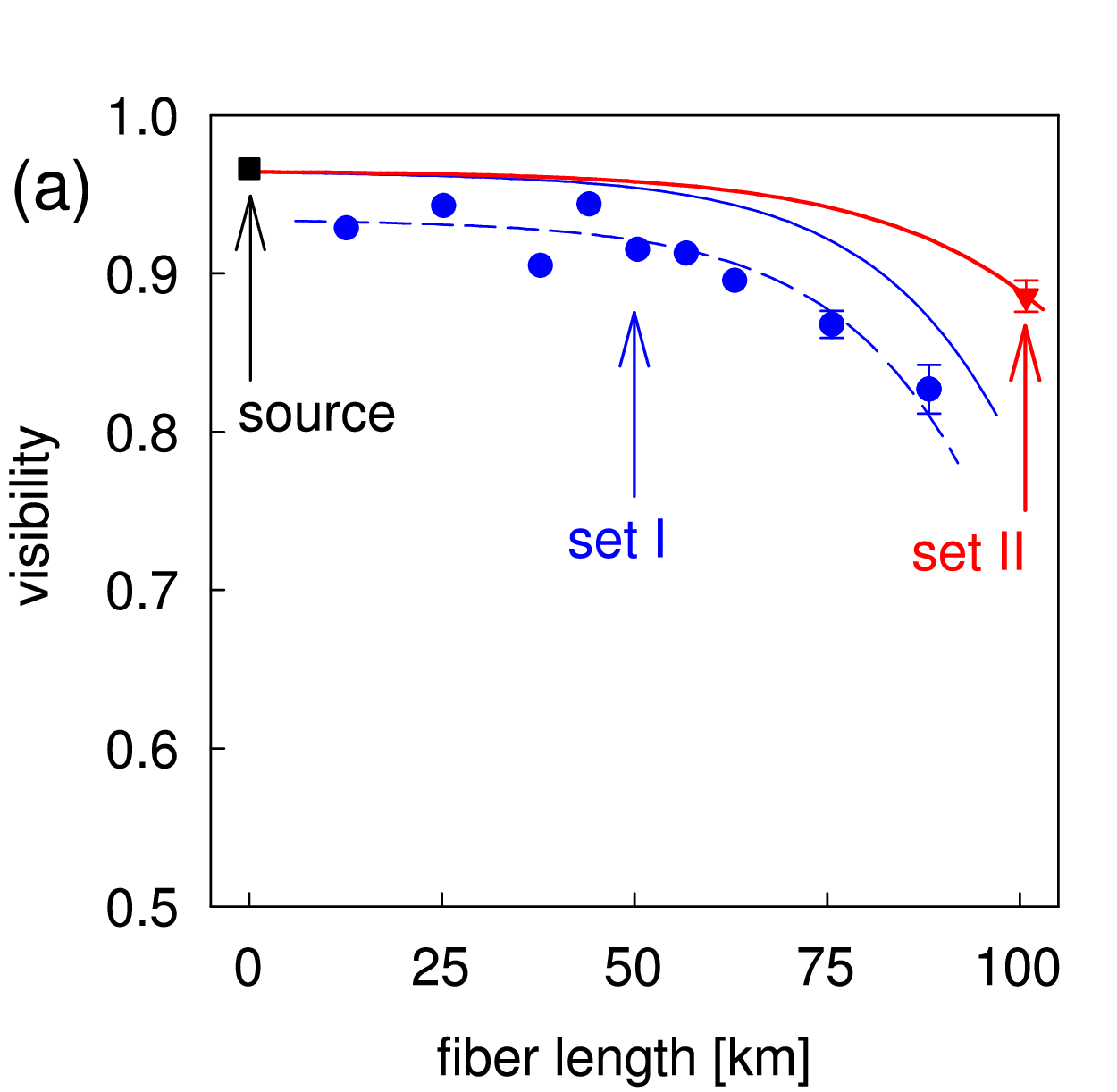}
\hfill
\includegraphics[height=6.0cm]{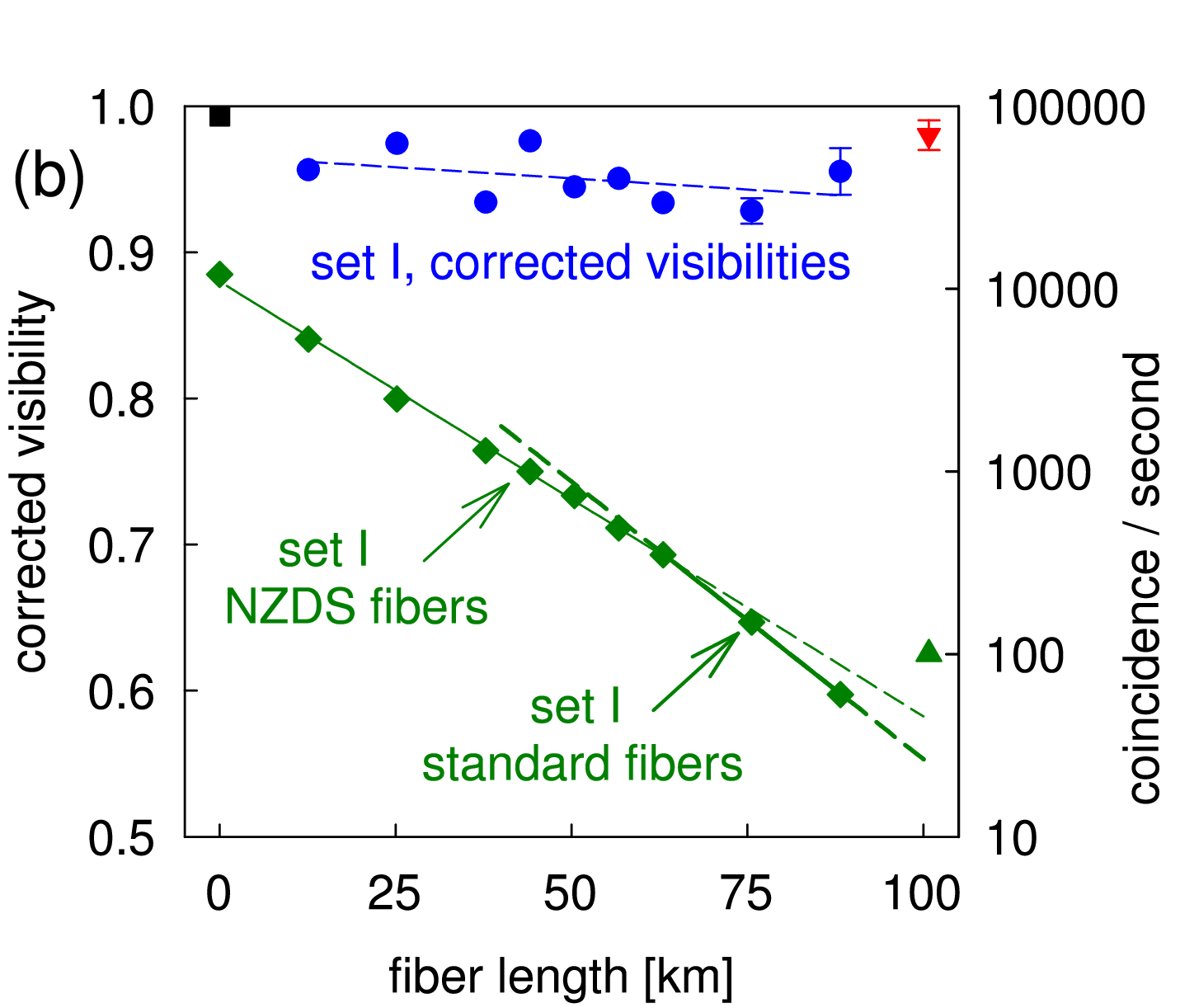}
\caption{(color online) (a) Uncorrected visibilities at the source
({\scriptsize$\blacksquare$}), in fiber set I
({\color{blue}$\bullet$}) and II~({\color{red}$\blacktriangledown$})
measured at different fiber lengths. The solid curves represent
model calculations (Eq.3) for set I and II. For the measured data of
set I we observe a discrepancy of 3 to 5\% (dashed curve) with the
model, indicating additional depolarization in the fiber.
(b)~Corrected visibilities (background subtracted) at the source
({\scriptsize$\blacksquare$}), in fiber set I
({\color{blue}$\bullet$}) and set II
({\color{red}$\blacktriangledown$}). Again, the fitted dashed line
to set I indicates a decrease in visibility from 4\% to 6\% below
the expected value. The measured coincidence count rates for set I
({\color{DarkGreen}$\blacklozenge$}) and II
({\color{DarkGreen}$\blacktriangle$}) are displayed on the right
scale. The increased loss due to higher chromatic dispersion in the
standard fibers ($>63$~km) is clearly visible in the different
slopes before and after 63 km.} \label{rawvis}
\end{center}
\end{figure}

The measured data points in Fig.~\ref{rawvis}(a) show the
uncorrected visibilities (no background subtracted) of the
\emph{H}\emph{V} and the $+-$ basis as a function of fiber length.
We display the average of the two visibilities since any negative
effect on the polarization induced by the fiber should affect both
polarizations basis equally. Indeed, we found the difference of the
two visibilities (\emph{H}\emph{V}, $+-$) to be less than 3\%.

After the first spool, a decrease in visibility from 96.6\% at the
source to 93\% is observed. The visibility remains approximately
this value for transmission distances up to 60~km before decreasing
steadily to 82.7\% at 88~km. For the 101~km fiber we see yet a
higher value of 88.6\%, because of the larger number of coincidences
and hence greater ratio between coincidences and dark counts. The
fluctuation seen in the data may in part be due to experimental
error in compensating for the polarization drift or depolarizing
effects in the fiber, like PMD. Error bars indicate the variance due
to Poissonian statistics in the count rates.
The upper plot in Fig.~\ref{rawvis}(b) shows the corrected
visibility ($V_{cor}$), the same data as in Fig.~\ref{rawvis}(a) but
with background counts removed. At the source (0~km), $V_{cor}$ was
measured to be 99.3\%. For all distances measured with fiber set~I,
$V_{cor}$ remains around 95\%. Using the 101~km fiber we see an
increase from 89\% (uncorrected) to $V_{cor}=98\%$.

Since such a simple background subtraction substantially improves
the correlation for far distant data points, we therefore
constructed a model to predict the reduction in the raw visibility
$V$ over the transmission length $l$ based on dark counts,
multi-pair emission and chromatic dispersion:
\eq%
V(l)=
\frac{(Max_{\,0}-Min_{\,0})\,T(l)\,F(l)}{((Max_{\,0}+Min_{\,0})\,F(l)+2\,R_{acc})\,T(l)+2\,R_{dark}}\label{vis_final}%
\eeq%
where $Max_{\,0}$ and $Min_{\,0}$ are the coincidence rates at 0~km,
$T(l)$ is the measured transmission after length $l$, $R_{acc}$ is
the accidental coincidence count rate at 0~km (due to multi-pairs)
and $R_{dark}$ is the detector dark count rate. $T(l)^2$ terms for
multi-pairs were omitted since their effect is less than 0.5\% on
the visibility. The dark count rate was measured by closing the
fiber input port of the InGaAs detector but still triggering it with
the Si-APD. Since $R_{dark}$ is independent with transmission
distance it will be the dominant factor in decreasing the visibility
at large distances. $F(l)$ describes the effect of chromatic
dispersion causing a broadening of the photon wavepacket which may
eventually become wider than the detector gate window, leading to a
decrease in $Max$ and $Min$. $F(l)$ is calculated from $\int D(t)\,
G(t,l) \, dt$, the overlap between the detector response (gate
window) $D(t)$ and the distance-dependent width of the gaussian
wavepacket $G(t,l)$. Note that $R_{acc}$ is not affected by
chromatic dispersion. The parameters $Max_{\,0}$, $Min_{\,0}$,
$R_{acc}$ and $R_{dark}$ were experimentally determined from
measurements of the source. $T(l)$ and $F(l)$ were found from
experimental characterization of the fibers.

All parameters in Eq.~\ref{vis_final} were measured independently
and the resulting curves for fiber set I and II are drawn in
Fig.~\ref{rawvis}(a). The curves start at the measured visibility at
the source, remaining constant up to about 60 km and then drop off
due to detector dark counts. The prediction for fiber set II shows a
slower decline due to higher transmission ($T(l)$) and lower
chromatic dispersion ($F(l)$).

The comparison of the experimental data to our basic model leads to
the conclusion that the decrease of visibility at larger distances
can be explained predominantly by detector dark counts and chromatic
dispersion effects and only a minor component is attributed to PMD
of the optical fiber:

\begin{itemize}

\item The model predicts the measured point very
accurately for set II, Fig.~\ref{rawvis}(a), implying that with this
fiber there are no additional effects leading to a reduction of the
visibility.

\item For set I, the measured raw visibility, lies on
average 3\%-5\% lower than the model curve. This difference
indicates that depolarization effects are present in the fiber which
are not covered by our model. However, the dashed guide-to-the-eye
(offset with the model curve) implies almost no additional loss of
coherence at longer lengths.

\item This tendency can also be seen when studying the corrected
visibilities in Fig.~\ref{rawvis}(b). After an initial drop in the
corrected visibility to 96\% (3\% lower than at the source), only a
small additional decrease of about 2\% for the remaining 75~km of
set I is observed. Similarly, for set II, a decrease of about 1.3\%
is observed for the whole 101~km transmission, giving an estimate of
how much the fiber disturbs the polarization state.

\item The stronger decrease for set I at shorter distances could be because the first
spool of fiber might have been slightly damaged or these fibers have
an exceptionally high PMD. Since PMD could not be directly measured
we did not include its effects in our model. We do however think
that this point still deserves further investigations and are
planning further experiments using short polarization maintaining
fibers to study the effects of PMD alone.

\end{itemize}


With the measured data, the performance of such a system for quantum
key distribution is estimated as follows: Due to lower $QE$ and a
lossy polarization measurement at Bob, the local coincidence rate
was decreased to $\sim$12 kcounts/s. The measured coincidence rates
are displayed in Fig.~\ref{rawvis}(b). The exponential decrease
corresponds to a loss of 0.23 dB/km on average for set I and a
smaller loss of 0.21 dB/km for set II. With the latter, a
coincidence rate of 104~counts/s could still be detected after
101~km. The same sifted key rate is expected for QKD~\cite{BBM92,
Poppe04}, if four detectors are used. An average qubit error rate of
5.7\% is estimated from the raw visibility. Ideal Entanglement based
systems have been proven secure against individual attacks if the
emission of multi-pairs is neglected~\cite{waks}. Under these
circumstances the privacy amplification results for the BB84
protocol can be applied. Including realistic error
correction~\cite{cascade} and privacy amplification for individual
attacks~\cite{QKD_Lut}, a secure key rate of 35 bit/s could
potentially be extracted after 101~km.

Finally, we can give an estimate from our model on the maximal
distance for distributing entanglement in fibers. For this we assume
a source with the same characteristics as presented here and a
detector at Bob's side with a lower dark count rate together with
the estimated 2\% loss of visibility for every 100~km of fiber. We
can neglect dispersion effects in this calculation ($F(l)=1$) since
smaller bandwidths can be achieved using longer
crystals~\cite{sebo}. Superconducting single-photon detectors (SSPD)
are currently the best choice for low noise detection in the telecom
band. SSPDs have already been used for detection of entangled
photons and exhibit a dark count rate of about 50~c/s (ungated) and
a $QE$ of 0.5\%~\cite{kumarsspd}. This would correspond to a total
dark count rate of 0.0015 c/s for our experimental set-up and would
allow the detection of quantum correlations without background
subtraction up to 200~km. The total rate at this distance would
however be in the order of one count in 100 seconds.

\section{Conclusion}
We have shown quantum polarization correlation between photon pairs
after 101~km of optical fiber transmission without background
subtraction, sufficient for secure QKD using commercially available
fibers and detectors. Through measurements of the two-photon
visibility and quantum state tomography we have observed only minor
depolarization of the state caused by PMD of the optical fiber.
Instead, the degree of observed quantum correlation is limited by
detector dark counts, multi-pair emissions and chromatic dispersion.
Modern fibers have a far smaller influence on the polarization state
than previously thought, even at distances of up to 101 km and
probably beyond this length.

We demonstrate that polarization encoding and this distribution
method can readily be implemented as a QKD scheme. Moreover,
polarization encoding provides more convenient control in
photon-atom interactions and together with its observed robustness
during optical fiber transmission makes polarization encoded qubits
ideal for future long-distance quantum communication applications,
like quantum repeater networks.

\setcounter{secnumdepth}{0}
\section{Acknowledgments}
We would like to thank Sebastien Sauge and Daniel Ljunggren (KTH,
Kista) for the intense and ongoing discussions on generation and
distribution of entanglement. In addition we like to thank Reinhard
Binder from Mattig-Schauer and David Peckham from OFS for the loan
of the 101 km of TrueWave\copyright \hspace{0.5mm} optical fiber. We
also acknowledge the financial support from Austrian Research
Centers GmbH, Austrian Science Fund FWF (SFB15) and Stadt Wien. This
work was supported by the European Commission through the integrated
projects SECOQC (Contract No. IST-2003-506813) and QAP (No. 015846).


\end{document}